\documentclass[twocolumn,english,aps,prl,showpacs,superscriptaddress]{revtex4-1}
\usepackage[T1]{fontenc}
\usepackage[latin9]{inputenc}
\usepackage{color}
\usepackage{array}
\usepackage{units}
\usepackage{amsmath}
\usepackage{amssymb}
\usepackage{graphicx}

\makeatletter

\providecommand{\tabularnewline}{\\}

\usepackage{babel}
\usepackage{dsfont}

\usepackage{hyperref}
\usepackage{times}
\def\equationautorefname~#1\null{Equation (#1)\null}

\makeatother

\usepackage{babel}
\begin{document}
\title{Quantum dynamics of collective spin states in a thermal gas}
\author{Roy Shaham}
\email{roy.shaham@weizmann.ac.il}

\affiliation{Department of Physics of Complex Systems, Weizmann Institute of Science,
Rehovot 76100, Israel}
\affiliation{Rafael Ltd, IL-31021 Haifa, Israel}
\author{Or Katz}
\affiliation{Department of Physics of Complex Systems, Weizmann Institute of Science,
Rehovot 76100, Israel}
\affiliation{Rafael Ltd, IL-31021 Haifa, Israel}
\author{Ofer Firstenberg}
\affiliation{Department of Physics of Complex Systems, Weizmann Institute of Science,
Rehovot 76100, Israel}
\begin{abstract}
Ensembles of alkali-metal or noble-gas atoms at room temperature and
above are widely applied in quantum optics and metrology owing to
their long-lived spins. Their collective spin states maintain nonclassical
nonlocal correlations, despite the atomic thermal motion in the bulk
and at the boundaries. Here we present a stochastic, fully quantum
description of the effect of atomic diffusion in these systems. We
employ the Bloch-Heisenberg-Langevin formalism to account for the
quantum noise originating from diffusion and from various boundary
conditions corresponding to typical wall coatings, thus modeling the
dynamics of nonclassical spin states with spatial interatomic correlations.
As examples, we apply the model to calculate spin noise spectroscopy,
temporal relaxation of squeezed spin states, and the coherent coupling
between two spin species in a hybrid system.
\end{abstract}
\maketitle

\section{Introduction\label{sec:Introduction}}

Gaseous spin ensembles operating at room temperature and above have
attracted much interest for decades. At ambient conditions, alkali-metal
vapors and odd isotopes of noble gases exhibit long spin-coherence
times, ranging from milliseconds to hours \citep{Happer1972OPIntro,Happer1977SERF,katz2018storage1sec,balabas2010minutecoating,walker1997SEOPReview,Walker2017He3review}.
These spin ensembles, consisting of a macroscopic number of atoms,
are beneficial for precision sensing, searches of new physics, and
demonstration of macroscopic quantum effects \citep{Happer2010book,Brown2010RomalisCPTviolation,Sheng2013RomalisSubFemtoTesla,Budker2007OpticalMagnetometryI,Budker2013OpticalMagnetometryII,Crooker2004SNSmagnetometerNature,ItayAxions2019arxiv}.
In particular, manipulations of collective spin states allow for demonstrations
of basic quantum phenomena, including entanglement, squeezing, and
teleportation \citep{Julsgaard2001PolzikEntanglement,Sherson2006PolzikTeleportationDemo,Jensen2011PolzikSqueezingStorage,Polzik2010ReviewRMP}
as well as storage and generation of photons \citep{Eisaman2005LukinSinglePhoton,Peyronel2012LukinInteractingSinglePhotons,Gorshkov2011LukinRydbergBlockadePhotonInteractions,Borregaard2016SinglePhotonsOnMotionallyAveragedMicrocellsNcomm}.
It is the collectively enhanced coupling and the relatively low noise
offered by these spin ensembles that make them particularly suitable
for metrology and quantum information applications.

Thermal atomic motion is an intrinsic property of the dynamics in
gaseous systems. Gas-phase atoms, in low-pressure room-temperature
systems, move at hundreds of meters per second in ballistic trajectories,
crossing the cell at sub-millisecond timescales and interacting with
its boundaries. To suppress wall collisions, buffer gas is often introduced,
which renders the atomic motion diffusive via velocity-changing collisions
\citep{Kastler1957OPreview}. At the theory level, the effect of diffusion
on the mean spin has been extensively addressed, essentially by describing
the evolution of an impure (mixed) spin state in the cell using a
mean-field approximation \citep{MasnouSeeuwsBouchiat1967diffusion,WuHapper1988CoherentCellWallTheory,Li2011RomalisMultipass,Firstenberg2007DickeNarrowingPRA,Firstenberg2010CoherentDiffusion,XiaoNovikova2006DiffusionRamsey}.
This common formalism treats the spatial dynamics of an average atom
in any given position using a spatially dependent density matrix.
It accurately captures the single-atom dynamics but neglects both
interatomic correlations and thermal fluctuations associated with
the spin motion and collisions.

Nonclassical phenomena involving collective spin states, such as transfer
of quantum correlations between nonoverlapping light beams by atomic
motion \citep{Xiao2019MultiplexingSqueezedLightDiffusion,Bao2019SpinSqueezing,bao2016spinSqueezing},
call for a quantum description of the thermal motion. For spin-exchange
collisions, which are an outcome of thermal motion, such a quantum
description has received much recent attention \citep{Kong2018MitchellAlkaliSEEntanglement,weakcollisions2019arxiv,AlkaliNobleEntanglementKatz2020PRL,Dellis2014SESpinNoisePRA,Mouloudakis2019SEwavefunctionUnraveling,Mouloudakis2020SEbipartiteEntanglement,Vasilakis2011RomalisBackactionEvation,Roy2015SNScrosscorrelations}.
However, the more direct consequences of thermal motion, namely, the
stochasticity of the spatial dynamics in the bulk and at the system's
boundaries, still lack a proper fully quantum description.

In this paper, we describe the effect of spatial diffusion on the
quantum state of warm spin gases. Using the Bloch-Heisenberg-Langevin
formalism, we identify the dissipation and noise associated with atomic
thermal motion and with the scattering off the cell boundaries. Existing
significant work in this field relies primarily on mean-field models,
which address both wall coupling \citep{seltzer2009RomalishighTcoating}
and diffusion in unconfined systems \citep{Lucivero2017RomalisDiffusionCorrelation}.
The latter work derives the correlation function of diffusion-induced
quantum noise from the correlation function of mass diffusion in unconfined
systems. Here we derive the quantum noise straight out of Brownian
motion considerations and provide a solution for confined geometries.
Our model generalizes the mean-field results and enables the description
of interatomic correlations and collective quantum states of the ensemble.
We apply the model to highly polarized spin vapor and analyze the
effect of diffusion in various conditions, including spin noise (SN)
spectroscopy \citep{Sinitsyn2016SpinNoiseSpectroscopySNSreview,Katsoprinakis2007SpinNoiseRelaxation,Crooker2004SNSmagnetometerNature,Crooker2014PRLspectroscopy,Lucivero2016MitchellSpinNoiseSpectrscopySqueezedLight,Lucivero2017MitchellNoiseSpectroscopyFundumentals},
spin squeezing \citep{Kong2018MitchellAlkaliSEEntanglement,Julsgaard2001PolzikEntanglement},
and coupling of alkali-metal to noble-gas spins in the strong-coupling
regime \citep{weakcollisions2019arxiv,AlkaliNobleEntanglementKatz2020PRL}.

The paper is arranged as follows. We derive in Sec.~\ref{sec:Model}
the Bloch-Heisenberg-Langevin model for the evolution of the collective
spin operator due to atomic Brownian motion and cell boundaries. We
focus on highly polarized ensembles in Sec.~\ref{sec:Polarized-ensemb}
and provide the model solutions. In Sec.~\ref{sec:Applications},
we present several applications of our model. We discuss how it is
employed to describe the temporal evolution, to calculate experimental
results, to provide insight, and to optimize setups for specific tasks.
Limits of our model and differences from existing models, as well
as future prospects, are discussed in Sec.~\ref{sec:Discussion}.
We provide appendices that elaborate on the quantum noise produced
by thermal motion (Appendix \ref{sec:diffusion-noise}), a simplified
model for analyzing the scattering off the cell walls (Appendix \ref{sec:wall-coupling-toy}),
means of solving the Bloch-Heisenberg-Langevin equation (Appendix
\ref{sec:solution-of-diffusion-relaxation}), and the Faraday rotation
scheme used herein (Appendix \ref{sec:Faraday-rotation-measurement}).

\begin{figure}
\includegraphics[width=1\columnwidth]{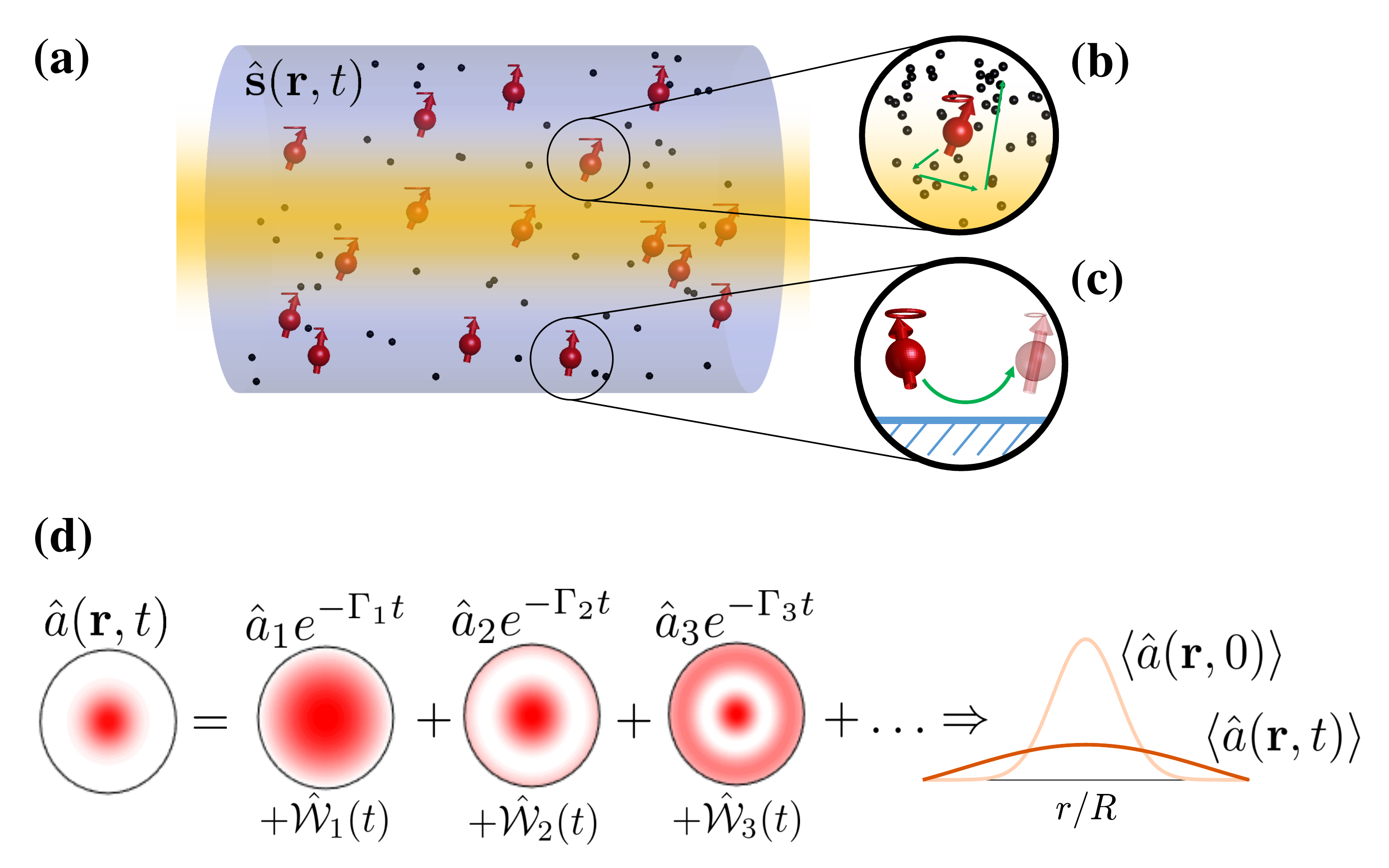}

\caption{(a) Atomic spins in the gas phase, comprising a collective quantum
spin $\mathbf{\hat{s}}(\mathbf{r},t)$ and undergoing thermal motion.
(b) In the diffusive regime, the spins spatially redistribute via
frequent velocity-changing collisions. (c) Collisions (local interaction)
with the walls of the gas cell may fully or partially depolarize the
spin state. (d) Diffusion and wall collisions lead to a multimode
evolution, here exemplified for a spin excitation $\hat{a}(\mathbf{r},t)\propto\hat{\mathrm{s}}_{x}(\mathbf{r},t)-i\hat{\mathrm{s}}_{y}(\mathbf{r},t)$
with an initial Gaussian-like spatial distribution $\langle\hat{a}(\mathbf{r},t)\rangle$
and for destructive wall collisions. In addition to the mode-specific
decay $\Gamma_{n}$, each spatial mode accumulates mode-specific quantum
noise $\mathcal{\hat{W}}_{n}(t)$.\label{fig:diffusion-illustration}}
\end{figure}

\section{Model\label{sec:Model}}

Consider a warm ensemble of $N_{\mathrm{a}}$ atomic spins confined
in a cell, as illustrated in Fig.~\ref{fig:diffusion-illustration}a.
Let $\mathbf{r}_{a}(t)$ be the classical location of the $a$th atom
at time $t$ and define the single-body density function at some location
$\mathbf{r}$ as $n_{a}(\mathbf{r})=\delta[\mathbf{r}-\mathbf{r}_{a}(t)]$.
We denote the spin operator of the $a$th atom by $\mathbf{\hat{s}}_{a}$
and define the space-dependent collective spin operator as $\mathbf{\hat{s}}(\mathbf{r},t)=\sum_{a=1}^{N_{\mathrm{a}}}\mathbf{\hat{s}}_{a}n_{a}(\mathbf{r})$.
While formally $\mathbf{\hat{s}}(\mathbf{r},t)$ is sparse and spiked,
practical experiments address only its coarse-grained properties,
e.g. ,due to finite spatial scale of the employed optical or magnetic
fields. The time evolution of the collective spin operator is given
by 
\begin{equation}
\frac{\partial\hat{\mathbf{s}}}{\partial t}=\sum_{a=1}^{N_{\mathrm{a}}}\frac{\partial\hat{\mathbf{s}}_{a}}{\partial t}n_{a}+\hat{\mathbf{s}}_{a}\frac{\partial n_{a}}{\partial t}.\label{eq:spin-equation}
\end{equation}
Here the first term accounts for the internal degrees of freedom,
including the local Hamiltonian evolution of the spins and spin-spin
interactions, while the second term accounts for the external degrees
of freedom, namely, for motional effects. The focus of this paper
is on the second term, considered in the diffusion regime as illustrated
in Fig.$\,$\ref{fig:diffusion-illustration}b. We consider the first
term only for its contribution to the boundary conditions, via the
effect of wall collisions as illustrated in Fig.$\,$\ref{fig:diffusion-illustration}c.
In the following, we first derive the equations governing the quantum
operator $\mathbf{\hat{s}}(\mathbf{r},t)$ in the bulk and subsequently
introduce the effect of the boundaries.

\subsection{Diffusion in the bulk\label{subsec:Thermal-motion}}

We consider the limit of gas-phase atoms experiencing frequent, spin-preserving,
velocity-changing collisions, such as those characterizing a dilute
alkali-metal vapor in an inert buffer gas. In this so-called Fickian
diffusion regime, the atomic motion is diffusive, and the local density
evolution can be described by the stochastic differential equation
\citep{dean1996langevinDiffusion} 
\begin{equation}
\partial n_{a}/\partial t=D\nabla^{2}n_{a}+\boldsymbol{\nabla}(\boldsymbol{\eta}\sqrt{n_{a}}),\label{eq:Dean-diffusion}
\end{equation}
where $D$ is the diffusion coefficient, and $\boldsymbol{\eta}$
is a white Gaussian stochastic process the components of which satisfy
$\langle\eta_{i}(\mathbf{r},t)\eta_{j}(\mathbf{r}',t')\rangle_{\mathrm{c}}=2D\delta_{ij}\delta(\mathbf{r}-\mathbf{r}')\delta(t-t')$
for $i,j=x,y,z$. We use $\langle\cdot\rangle_{\mathrm{c}}$ to represent
ensemble average over the classical atomic trajectories, differing
from the quantum expectation value $\langle\cdot\rangle$. The first
term in Eq.$\,$(\ref{eq:Dean-diffusion}) leads to delocalization
of the atomic position via deterministic diffusion, while the second
term introduces fluctuations that localize the atoms to discrete positions.
Equation (\ref{eq:Dean-diffusion}), derived by Dean for Brownian
motion in the absence of long-range interactions \citep{dean1996langevinDiffusion},
is valid under the coarse-grain approximation, when the temporal and
spatial resolutions are coarser than the mean-free time and path between
collisions.

Substituting $\partial n_{a}/\partial t$ into Eq.$\,$(\ref{eq:spin-equation}),
we obtain the Bloch-Heisenberg-Langevin dynamical equation for the
collective spin 
\begin{equation}
\partial\hat{\mathbf{s}}/\partial t=i[\mathcal{H},\hat{\mathbf{s}}]+D\nabla^{2}\mathbf{\hat{s}}+\hat{\boldsymbol{f}}.\label{eq:spin-diffusion-equation}
\end{equation}
Here $\mathcal{H}$ is the spin Hamiltonian in the absence of atomic
motion, originating from the $\partial\hat{\mathbf{s}}_{a}/\partial t$
term in Eq.$\,$(\ref{eq:spin-equation}). The quantum noise operator
$\hat{\boldsymbol{f}}=\hat{\boldsymbol{f}}(\mathbf{r},t)$ is associated
with the local fluctuations of the atomic positions. It can be formally
written as $\hat{f}_{\mu}=\boldsymbol{\nabla}(\hat{\mathrm{s}}_{\mu}\boldsymbol{\eta}\slash\sqrt{n})$,
where $\mu=x,y,z$, and $n=\sum_{a}n_{a}$ is the atomic density.
The noise term has an important role in preserving the mean spin moments
of the ensemble. The commutation relation of different instances of
the noise $\hat{f}_{\mu}=\hat{f}_{\mu}(\mathbf{r},t)$ and $\hat{f}_{\nu}'=\hat{f}_{\nu}(\mathbf{r}',t')$
satisfies 
\begin{equation}
\langle[\hat{f}_{\mu},\hat{f}_{\nu}']\rangle_{\mathrm{c}}=2i\epsilon_{\xi\mu\nu}D(\boldsymbol{\nabla}\boldsymbol{\nabla}')\hat{\mathrm{s}}_{\xi}\delta(\mathbf{r}-\mathbf{r}')\delta(t-t'),\label{eq:diffusion-noise-commutation-relation}
\end{equation}
where $\epsilon_{\xi\mu\nu}$ is the Levi-Civita antisymmetric tensor.
These commutation relations ensure the conservation of spin commutation
relations $[\hat{\mathrm{s}}_{\mu}(\mathbf{r},t),\hat{\mathrm{s}}_{\nu}(\mathbf{r}',t)]=i\epsilon_{\xi\mu\nu}\hat{\mathrm{s}}_{\xi}\delta(\mathbf{r}-\mathbf{r}')$
on the operator level, compensating for the diffusion-induced decay
in the bulk due to the $D\nabla^{2}$ term. We provide the full derivation
of $\hat{\boldsymbol{f}}$ and its properties in Appendix \ref{sec:diffusion-noise}.

The spin noise p\textcolor{black}{rocess is temporally white and spatially
colored, with higher noise content for shorter wavelengths. The increase
of noise at a fine-grain scale counteracts the diffusion term, which
decreases the spin variations faster }at smaller length scales; this
is a manifestation of the fluctuation-dissipation theorem. Finally,
as expected, ensemble averaging over the noise realizations leaves
only the diffusion term in the mean-field Bloch equation for the spin
$\partial\langle\mathbf{s}\rangle/\partial t=D\nabla^{2}\langle\mathbf{s}\rangle$,
where $\langle\mathbf{s}\rangle=\langle\hat{\mathbf{s}}(\mathbf{r},t)\rangle$
is the spin expectation value at a course-grained position $\mathrm{\boldsymbol{r}}$.

\subsection{Boundary conditions\label{subsec:Cell-walls}}

We now turn to derive the contribution of wall collision to the quantum
dynamics of the collective spin. When the atoms diffuse to the boundaries
of the cell, their spin interacts with the surface of the walls. This
interaction plays an important role in determining the depolarization
and decoherence times of the total spin \citep{Kastler1957OPreview,Happer2010book}
and may also induce frequency shifts \citep{Volk1979CoherentWalls,Kwon1981CoherentXe131Wall,simpson1978CellWallNMR,Wu1987CoherentCellWallExp,WuHapper1988CoherentCellWallTheory}.
Bare glass strongly depolarizes alkali-metal atoms, and magnetic impurities
in the glass affect the nuclear spin of noble-gas atoms. To attenuate
the depolarization at the walls, cells can be coated with spin-preserving
coatings such as paraffin \citep{Alexandrov2002LightInducedDesorptionParaffin,Graf2005paraffinPRA,balabas2010minutecoating}
or octadecyltrichlorosilane (OTS) \citep{seltzer2009RomalishighTcoating}
for alkali-metal vapor and Surfasil or SolGel \citep{Driehuys1993HapperSurfasilPRA,Driehuys1995HapperSurfasilPRL,Breeze1999Surfasil,Hsu2000SolGelCoating}
for spin-polarized xenon. The coupling between the spins and the cell
walls constitutes the formal boundary conditions of Eq.~(\ref{eq:spin-diffusion-equation}).

In the mean-field picture, the wall coupling can be described as a
local scatterer for the spin-density matrix $\rho$. In this picture,
assisted by kinetic gas theory, the boundary conditions can be written
as \citep{MasnouSeeuwsBouchiat1967diffusion} 
\begin{equation}
(1+\tfrac{2}{3}\lambda\hat{\mathbf{n}}\cdot\boldsymbol{\nabla})\rho=(1-\tfrac{2}{3}\lambda\hat{\mathbf{n}}\cdot\boldsymbol{\nabla})\mathcal{S}\rho,\label{eq:mean-field-boundary}
\end{equation}
where $\mathcal{S}$ is the wall scattering matrix. Here $\lambda$
denotes the mean free path of the atoms, related to the diffusion
coefficient via $D=\lambda\bar{v}/3$, where $\bar{v}$ is the mean
thermal velocity.

We adopt a similar perspective in order to derive the coupling of
the collective spin $\hat{\mathbf{s}}$ with the walls in the Bloch-Heisenberg-Langevin
formalism. In this formalism, the scattering off the walls introduces
not only decay, but also fluctuations. In the Markovian limit, when
each scattering event is short, its operation on a single spin becomes
a stochastic density matrix
\begin{equation}
\mathcal{S}\hat{\mathbf{s}}_{a}=e^{-1/N}\hat{\mathbf{s}}_{a}+\hat{\boldsymbol{w}}_{a}.\label{eq:wall-scattering}
\end{equation}
Here $N$ denotes the average number of wall collisions a spin withstands
before depolarizing \citep{seltzer2009RomalishighTcoating}. The accompanied
quantum noise process is $\hat{\boldsymbol{w}}_{a}$; it ensures the
conservation of spin commutation relations at the boundary. 

Using the stochastic scattering matrix, we generalize the mean-field
boundary condition {[}Eq.~(\ref{eq:mean-field-boundary}){]} for
collective spin operators as
\begin{equation}
(1-e^{-1/N})\hat{\mathbf{s}}+\tfrac{2}{3}\lambda(1+e^{-1/N})(\hat{\mathbf{n}}\cdot\boldsymbol{\nabla})\hat{\mathbf{s}}=\hat{\boldsymbol{w}}.\label{eq:boundary-condition}
\end{equation}
Here $\hat{\boldsymbol{w}}(\mathbf{r},t)=\sum_{a}\hat{\boldsymbol{w}}_{a}n_{a}$,\textcolor{red}{{}
}for positions $\mathbf{r}$ on the cell boundary, is the collective
wall-coupling noise process affecting the local spin on the wall.
$\hat{\boldsymbol{w}}$ is zero on average and its statistical properties,
together with the derivation of Eq.$\,$(\ref{eq:wall-scattering}),
are discussed in Appendix \ref{sec:wall-coupling-toy}. The first
term in Eq.~(\ref{eq:boundary-condition}) describes the fractional
depolarization by the walls, and the second term describes the difference
between the spin flux entering and exiting the wall. If the wall coupling
also includes a coherent frequency-shift component, it can be appropriately
added to these terms. The term on the right-hand side describes the
associated white fluctuations.

In the limit of perfect spin-preserving coating, the boundary condition
becomes a no-flux (Neumann) condition satisfying $(\hat{\mathbf{n}}\cdot\boldsymbol{\nabla})\hat{\mathbf{s}}=0$,
and depolarization is minimized. This limit is realized for $N\gg R/\lambda$,
where $R$ is the dimension of the cell$\,$\footnote{For long-lived solutions of the diffusion equation, the flux towards
the wall $\hat{\mathbf{n}}\cdot\boldsymbol{\nabla}\hat{\mathbf{s}}$
is of the same order as $\hat{\mathbf{s}}/R$.} \citep{Happer2010book}. In the opposite limit of strongly depolarizing
walls, i.e., $N\lesssim1$, the (Dirichlet) boundary condition is
$\hat{\mathbf{s}}=\hat{\boldsymbol{w}}/(1-e^{-1/N})$ \footnote{This limit is obtained only when $\lambda\ll R$, which is also necessary
for the validity of the diffusion equation.}, rendering the scattered spin state random. For any other value of
$N$ (partially depolarizing walls), the boundary condition in Eq.~(\ref{eq:boundary-condition})
is identified as a stochastic Robin boundary condition \citep{ozisik2002boundarybook}.

\medskip{}

The two mechanisms discussed in this section --- the bulk diffusion
and the wall coupling --- are independent physical processes. This
is evident by the different parameters characterizing them --- $D$
and $N$ --- which are dictated by different physical scales, such
as buffer gas pressure and the quality of the wall coating. These
processes are different in nature; while wall coupling leads to spin
depolarization and thermalization, diffusion leads to spin redistribution
while conserving the total spin. They introduce independent fluctuations
and dissipation, and they affect the spins at different spatial domains
(the bulk and the boundary). That being said, both processes are necessary
to describe the complete spin dynamics in a confined volume, simultaneously
satisfying Eqs.$\,$(\ref{eq:spin-diffusion-equation}) and (\ref{eq:boundary-condition}).

\section{Polarized ensembles\label{sec:Polarized-ensemb}}

When discussing nonclassical spin states for typical applications,
it is beneficial to consider the prevailing limit of highly polarized
ensembles. Let us assume that most of the spins point downwards ($-\hat{\mathbf{z}}$).
In this limit, we follow the Holstein-Primakoff transformation \citep{Holstein1940PrimakoffHP,kittel1987forHPapprox}
and approximate the longitudinal spin component by its mean value
$\mathbf{\hat{\mathrm{s}}}_{z}(\mathbf{r},t)=\mathbf{\mathrm{s}}_{z}$
(with $\mathrm{s}_{z}=-n/2$ for spin 1/2). The ladder operator $\hat{\mathrm{s}}_{-}=\hat{\mathrm{s}}_{x}-i\hat{\mathrm{s}}_{y}$,
which flips a single spin downwards at position $\mathbf{r}$, can
be represented by the annihilation operator $\hat{a}=\hat{\mathrm{s}}_{-}/\sqrt{2|\mathrm{s}_{z}|}$.
This operator satisfies the bosonic commutation relations $[\hat{a}(\mathbf{r},t),\hat{a}^{\dagger}(\mathbf{r}',t)]=\delta(\mathbf{r}-\mathbf{r}')$.
Under these transformations, Eqs.~(\ref{eq:spin-diffusion-equation})
and (\ref{eq:boundary-condition}) become

\begin{align}
\partial\hat{a}/\partial t & =i[\mathcal{H},\hat{a}]+D\nabla^{2}\hat{a}+\hat{f},\label{eq:HP heisenberg langevin}\\
(1-e^{-1/N})\hat{a} & =-\tfrac{2}{3}\lambda(1+e^{-1/N})\hat{\mathbf{n}}\cdot\boldsymbol{\nabla}\hat{a}+\hat{w},\label{eq:HP boundary condition}
\end{align}
where both $\hat{f}=(\hat{f}_{x}-i\hat{f}_{y})/\sqrt{2|\mathbf{\mathrm{s}}_{z}|}$
and $\hat{w}=(\hat{w}_{x}+i\hat{w}_{y})/\sqrt{2|\mathbf{\mathrm{s}}_{z}|}$
are now vacuum noise processes (see Appendices \ref{sec:diffusion-noise}
and \ref{sec:wall-coupling-toy}; note that $\hat{f}$ is spatially
colored). Here, Eq.~(\ref{eq:HP heisenberg langevin}) describes
the spin dynamics in the bulk, while Eq.$\,$(\ref{eq:HP boundary condition})
holds at the boundary.

We solve Eqs.~(\ref{eq:HP heisenberg langevin}) and (\ref{eq:HP boundary condition})
by decomposing the operators into a superposition of nonlocal diffusion
modes $\hat{a}(\mathbf{r},t)=\sum_{n}\hat{a}_{n}(t)u_{n}(\mathbf{r})$.
We first identify the mode functions $u_{n}(\mathbf{r})$ by solving
the homogeneous Helmholtz equation $(D\nabla^{2}+\Gamma_{n})u_{n}(\mathbf{r})=0$,
where the eigenvalues $-\Gamma_{n}$ are fixed by the Robin boundary
condition {[}Eq.~(\ref{eq:HP boundary condition}) without the noise
term{]}. The operator $\hat{a}_{n}(t)=\int_{V}\hat{a}(\mathbf{r},t)u_{n}^{\ast}(\mathbf{r})d^{3}\mathbf{r}$,
where $V$ is the cell volume, annihilates a collective transverse
spin excitation with a nonlocal distribution $|u_{n}(\mathbf{r})|^{2}$
and a relaxation rate $\Gamma_{n}$. These operators satisfy the bosonic
commutation relation $[\hat{a}_{n},\hat{a}_{m}^{\dagger}]=\delta_{nm}$.
The noise terms $\hat{f}$ and $\hat{w}$ are decomposed using the
same mode-function basis. This leads to mode-specific noise terms
$\mathcal{\hat{W}}_{n}(t)$, operating as independent sources.

Assuming, for the sake of example, a magnetic (Zeeman) Hamiltonian
$\mathcal{H}=\omega_{0}\hat{\mathrm{s}}_{z}$, where $\omega_{0}$
is the Larmor precession frequency around a $\hat{\mathbf{z}}$ magnetic
field, the time evolution of the mode operators is given by
\begin{equation}
\hat{a}_{n}(t)=\hat{a}_{n}(0)e^{-(i\omega_{0}+\Gamma_{n})t}+\mathcal{\hat{W}}_{n}(t).\label{eq:eigenmode-time-evolution}
\end{equation}
The multimode decomposition and evolution are illustrated in Fig.$\,$\ref{fig:diffusion-illustration}d,
showing the first angular-symmetric mode distributions $u_{n}(\mathbf{r})$
of a cylindrical cell. In Table \ref{tab:Diffusion-mode-solutions},
we provide explicit solutions of the mode bases and associated decay
rates for any given boundary properties in either rectangular, cylindrical,
or spherical cells. The solution procedure and the corresponding decomposition
of the noise terms are demonstrated for an exemplary one-dimensional
(1D) geometry in Appendix \ref{sec:solution-of-diffusion-relaxation}.
We note that, asymptotically, the decay of high-order modes ($n\gg1$)
is independent of cell geometry and is approximately given by $\Gamma_{n}\sim D(\pi nV^{-1/3})^{2}$,
where $\pi nV^{-1/3}$ approximates the mode's wave number.

\section{Applications\label{sec:Applications}}

The outlined Bloch-Heisenberg-Langevin formalism applies to various
experimental configurations and applications. It should be particularly
useful when two constituents of the same system have different spatial
characteristics, leading to different spatial modes. That occurs,
for example, when coupling spins to optical fields {[}Fig.$\,$\ref{fig:diffusion-illustration}d{]}
or when mixing atomic species with different wall couplings. In this
section, we consider three such relevant, real-life cases.

\begin{figure*}
\includegraphics[width=1.9\columnwidth]{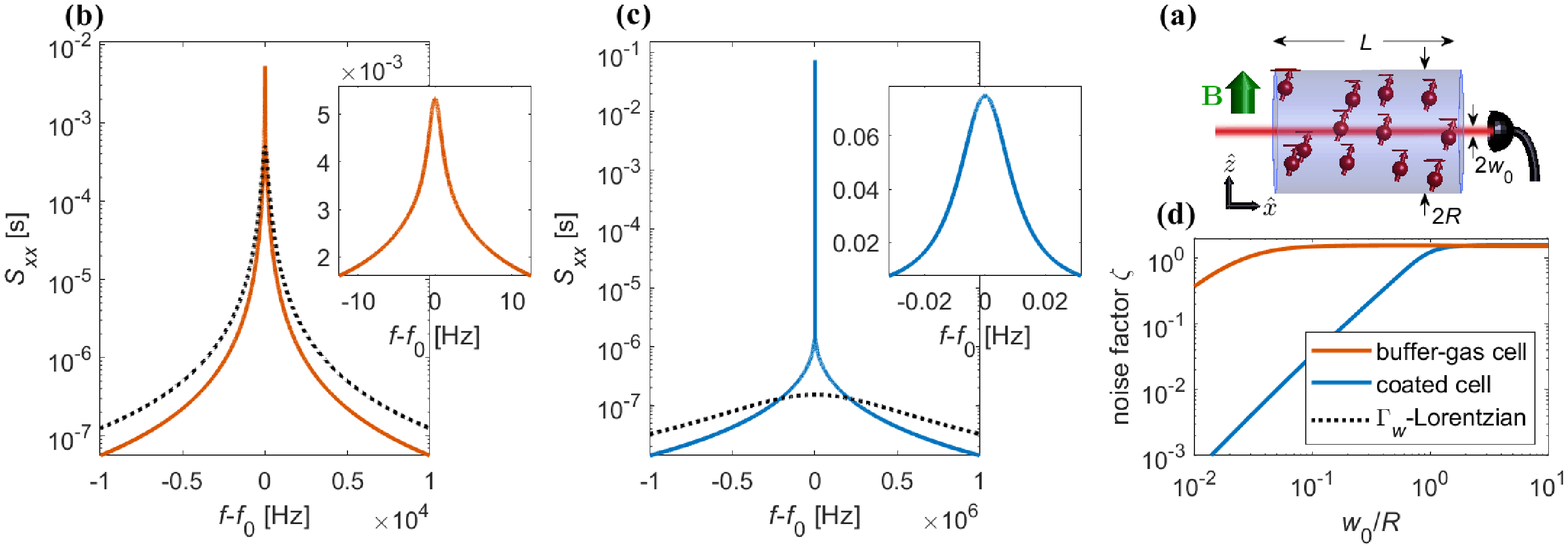}

\caption{Effect of thermal motion in spin noise spectroscopy. (a) The spins
are initially polarized along a magnetic field $B\hat{\mathbf{z}}$,
and the spin projection $\langle\hat{\mathrm{s}}_{x}\rangle$ is measured
by Faraday rotation using a nondiverging Gaussian probe beam. The
calculation assumes a probe waist radius of $w_{0}=\unit[1]{mm}$
and a cylindrical cell with radius $R=1$ cm and length $L=3$ cm.
(b) Spin noise spectrum for an uncoated cell with a dense buffer gas
($D=1\,\mathrm{cm^{2}/s}$, $N\protect\leq1$), calculated using Eq.$\,$(\ref{eq:APN-PSD}).
Many spatial (diffusion) modes contribute to the noise, and thus the
total signal is a weighted sum of varying Lorentzians, producing a
cusp profile. (c) Spin noise spectrum for a cell coated with high-quality
paraffin coating ($D=3\cdot10^{3}\,\mathrm{cm^{2}/s}$, $N=10^{6}$).
The cusp is wider (since $D$ is larger), except for an additional
sharp feature associated with the uniform spatial mode, the slow decay
of which is governed by wall collisions. The dotted lines in (b) and
(c) are a simple Lorentzian with width $\Gamma_{w}=\pi^{2}D/w_{0}^{2}$,
provided as reference for a single-mode approximation. (d) Noise content
in the vicinity of the resonance (as defined in the main text) for
the same two cells and varying probe waists. The narrower the probe
beam, the larger its overlap with the high-order, rapidly decaying,
diffusion modes, thus leading to weaker signal in the central resonance
feature. This effect becomes more pronounced for lower buffer gas
pressures.\label{fig:APN-PSD}}
\end{figure*}

\subsection{Spin noise spectroscopy\label{subsec:Noise-spectrum}}

SNS allows one to extract physical data out of the noise properties
of the spin system. It is used for magnetometry with atomic ensembles
in or out of equilibrium \citep{Crooker2004SNSmagnetometerNature,Crooker2014PRLspectroscopy,Lucivero2017MitchellNoiseSpectroscopyFundumentals,Katsoprinakis2007SpinNoiseRelaxation,Tang2020DiffusionLowPressure_PRA},
for low-field NMR \citep{Tayler2016zeroFieldNMR}, for fundamental
noise studies aimed at increasing metrological sensitivity \citep{Sheng2013RomalisSubFemtoTesla,Crooker2004SNSmagnetometerNature},
and more \citep{Sinitsyn2016SpinNoiseSpectroscopySNSreview}. SNS
is also used to quantify interatomic correlations in squeezed states,
when it is performed with precision surpassing the standard quantum
limit \citep{Julsgaard2001PolzikEntanglement,Katsoprinakis2007SpinNoiseRelaxation,Kong2018MitchellAlkaliSEEntanglement,Roy2015SNScrosscorrelations}.

Spin noise in an alkali-metal vapor is affected by various dephasing
mechanisms. Here we describe the effect of diffusion, given a spatially
fixed light beam employed to probe the spins. Since this probe beam
may overlap with several spatial modes of diffusion, the measured
noise spectrum would depend on the beam size, cell dimensions, and
diffusion characteristics. On the mean-field level, this effect has
been described by motion of atoms in and out of the beam \citep{Pugatch2009UniversalDiffusion,XiaoNovikova2006DiffusionRamsey}.
Here we calculate the SNS directly out of the quantum noise induced
by the thermal motion as derived above.

For concreteness, we consider two cylindrical cells of radius $R=1$
cm and length $L=3$ cm. One cell contains 100 Torr of buffer gas,
providing for $\lambda=\unit[0.5]{\mu m}$ and $D=1\,\mathrm{cm^{2}/s}$,
and no spin-preserving coating $N\lesssim1$ (e.g., as in Ref.~\citep{Kong2018MitchellAlkaliSEEntanglement}).
The other cell has a high-quality paraffin coating, allowing for $N=10^{6}$
wall collisions before depolarization \citep{balabas2010minutecoating},
and only dilute buffer gas originating from outgassing of the coating,
such that $\lambda=\unit[1]{mm}$ and $D=3\cdot10^{3}\,\mathrm{cm^{2}/s}$
rendering the atomic motion in the Fickian regime ($\lambda\ll R,L$)
\citep{sekiguchi2016JapaneseParaffinOutgassing,Hatakeyama2019ParaffinOutgassing2}.
A probe beam with waist radius $w_{0}$ measures the alkali-metal
spin $\hat{\mathbf{x}}$ component, oriented along the cylinder axis
as presented in Fig.$\,$\ref{fig:APN-PSD}a. The cell is placed inside
a magnetic field $\mathbf{B}=2\pi f_{0}/g_{\mathrm{a}}\cdot\hat{\mathbf{z}}$
pointing along the spin polarization, where $g_{\mathrm{a}}$ is the
alkali-metal gyromagnetic ratio.

In Appendix \ref{sec:Faraday-rotation-measurement}, we review the
measurement details and calculate the spin noise spectral density
$S_{xx}(f)$ for both cells 
\begin{equation}
S_{xx}(f)=\sum_{n}\frac{|I_{n}^{(\mathrm{G})}|^{2}}{4}\frac{2\tilde{P}\Gamma_{n}}{\Gamma_{n}^{2}+4\pi^{2}(f-f_{0})^{2}},\label{eq:APN-PSD}
\end{equation}
where $f$ is the frequency in which the SNS is examined, $\Gamma_{n}$
is again the decay rate of the $n$th diffusion mode, $I_{n}^{(\mathrm{G})}$
is the overlap of the Gaussian probe beam with that mode, and $\tilde{P}$
depends on the spin statistics and on the polarization, such that
$\tilde{P}=1$ for highly polarized ensembles and for spin 1/2.

The calculated spectra are shown in Figs.$\,$\ref{fig:APN-PSD}b
and \ref{fig:APN-PSD}c for $w_{0}=1$ mm. The cusp-like spectra originate
from a sum of Lorentzians, the relative weights of which correspond
to the overlap of the probe beam with each given mode $|I_{n}^{(\mathrm{G})}|^{2}$.
In the past, this cusp was identified as a universal phenomena \citep{Pugatch2009UniversalDiffusion},
while here we recreate this result using the eigenmodes and accounting
for the boundary. With spin-preserving coating, the uniform mode $n=0$
decays slower, and its contribution to the noise spectrum is much
more pronounced, while the higher-order modes decay faster due to
lack of buffer gas. 

The dominance of the central narrow feature thus depends on the overlap
of the probe with the least-decaying mode $|I_{0}^{(G)}|^{2}$. To
quantify it, we define the unitless noise content $\zeta=\int_{-f_{1/2}}^{f_{1/2}}S_{xx}(f)df$
as the fraction of the noise residing within the full width at half
maximum of the spectrum. Figure \ref{fig:APN-PSD}d shows $\zeta$
for different beam sizes $w_{0}/R$. Evidently, the spin resonance
is more significant in the buffer gas cell, unless the probe beam
covers the entire cell. This should be an important consideration
in the design of such experiments.

\subsection{Squeezed-state lifetime\label{subsec:Squeezed-state-lifetime}}

When the spin noise is measured with a sensitivity below the standard
quantum limit, the spin ensemble is projected into a collective squeezed
spin state. Such measurements are done primarily using optical Faraday
rotation in paraffin-coated cells \citep{Julsgaard2001PolzikEntanglement,Sherson2006PolzikTeleportationDemo,Wasilewski2010PolzikEntanglementMagnetometry,Jensen2011PolzikSqueezingStorage}
and recently also in the presence of buffer gas \citep{Kong2018MitchellAlkaliSEEntanglement}.
The duration of the probe pulse and the spatial profile of the probe
beam determine the spatial profile of the squeezed spin state and
hence its lifetime.

\begin{figure}
\includegraphics[width=0.95\columnwidth]{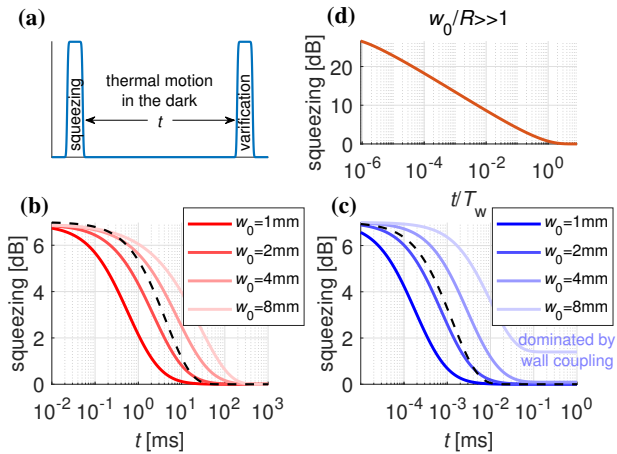}

\caption{Lifetime of spin squeezing. (a) Experimental sequence comprising a
short measurement (squeezing) pulse, followed by dephasing in the
dark due to thermal motion for duration $t$, and a verification pulse.
The same probe beam is used for both pulses. In the calculations,
the Gaussian distribution is initially squeezed by the measurement
to spin variance of $\langle\hat{x}_{\mathrm{G}}^{2}(0)\rangle=0.05$
(7-dB squeezing). We take the same cell geometry as in Fig.~\ref{fig:APN-PSD}
($R=1$ cm, $L=3$ cm). (b) Degree of spin squeezing vs time in the
buffer gas cell, calculated using Eq.$\,$(\ref{eq:squeezing-vs-time}).
Spin squeezing exhibits multiexponential decay associated with multiple
diffusion modes. Larger probe beams lead to longer squeezing lifetimes,
as the beam overlaps better with lower-order modes. (c) Spin squeezing
in the coated cell. In a coated cell, the decay rate of the uniform
diffusion mode, dominated by wall coupling, is substantially lower
than that of higher-order modes. Therefore, ensuring a significant
overlap of the probe beam with the uniform mode is even more important
in coated cells for maximizing the squeezing lifetime. The dotted
line in (b) and (c) is a single exponential decay $\langle\hat{x}^{2}(t)\rangle=\langle\hat{x}^{2}(0)\rangle e^{-2\Gamma_{w}t}+(1-e^{-2\Gamma_{w}t})/4$,
shown for reference with $\Gamma_{w}=\pi^{2}D/w_{0}^{2}$ and $w_{0}=4$
mm; note the difference in time-scales between (b) and (c). (d) An
extremely squeezed state relies more on higher-order spatial modes
and thus loses its squeezing degree rapidly (time is normalized by
$T_{w}=R^{2}/\pi^{2}D$). The calculation is initialized with a uniform
distribution of squeezing and includes the first 1000 radial modes,
required for convergence.\label{fig:multiexpnential-decay-of-squeezing}}
\end{figure}

We shall employ the same two cells and geometry from the previous
section {[}see Fig.$\,$\ref{fig:APN-PSD}a{]}. Given a probe pulse
duration much shorter than $w_{0}^{2}/D$ and assuming the measurement
sensitivity surpasses the standard quantum limit, a squeezed state
is formed, with initial spin variance $\langle\hat{x}_{\mathrm{G}}^{2}(0)\rangle\le1/4$,
where $\hat{x}_{G}(t)$ is the measured spin operator {[}defined in
Appendix \ref{sec:Faraday-rotation-measurement} as a weighted integral
over the local operator $\hat{x}(\mathbf{r},t)${]}. The state is
remeasured (validated) after some dephasing time $t$ {[}see Fig.$\,$\ref{fig:multiexpnential-decay-of-squeezing}a{]}.
In this type of experiments, narrow beams are often preferable, as
the local intensity affects the measurement sensitivity, and since
narrow beams simplify the use of optical cavities. Considering the
different diffusion modes $u_{n}(\mathbf{r})$ with their decay rates
$\Gamma_{n}$, we use Eq.$\,$(\ref{eq:eigenmode-time-evolution})
to calculate the evolution in the dark of the spin variance 
\begin{equation}
\langle\hat{x}_{\mathrm{G}}^{2}(t)\rangle=(\sum_{n}|I_{n}^{(G)}|^{2}e^{-\Gamma_{n}t})^{2}(\langle\hat{x}_{\mathrm{G}}^{2}(0)\rangle-1/4)+1/4.\label{eq:squeezing-vs-time}
\end{equation}

Figures \ref{fig:multiexpnential-decay-of-squeezing}b and \ref{fig:multiexpnential-decay-of-squeezing}c
present the calculated evolution. As expected, a narrow probe beam
squeezes the atoms with which it overlaps, which are spanned in a
superposition of diffusion modes {[}the first low-order modes in the
buffer gas cell are visualized in Fig.$\,$\ref{fig:diffusion-illustration}d{]},
leading to multimode temporal dynamics. The measured squeezing decreases
due to atoms diffusing out of the beam, as manifested by the exponential
decay of each spatial mode. The importance of thermal motion grows
as the degree of squeezing increases, as the latter relies on squeezing
in higher-spatial modes. To see this, we plot in Fig.~\ref{fig:multiexpnential-decay-of-squeezing}d
the decay of squeezing in the buffer gas cell with a wide probe beam
and with the initial state extremely squeezed $\langle\hat{x}^{2}(\mathbf{r},t=0)\rangle\lll1/4$.
The squeezing rapidly decays, as a power law, until only the lowest-order
mode remains squeezed. This indicates the practical difficulty in
achieving and maintaining a high degree of squeezing. An interesting
behavior is apparent for the case of a large beam in a coated cell
{[}Fig.~\ref{fig:multiexpnential-decay-of-squeezing}c, $w_{0}=8$
mm{]}. Here, the significant overlap with the uniform produces a certain
degree of squeezing that is especially long lived.

These results demonstrate the significance of accounting for many
diffusion modes when considering fragile nonclassical states or high-fidelity
operations. For example, the presented calculations for the 25-dB
squeezing require 1000 modes to converge.

\subsection{Coupling of alkali-metal spins to noble-gas spins \label{subsec:weak-collisions}}

Lastly, we consider collisional spin exchange between two atomic species
\citep{Katz2015SERFHybridization,Happer1977SERF,Dellis2014SESpinNoisePRA,Happer2010book,Mouloudakis2019SEwavefunctionUnraveling,weakcollisions2019arxiv,Roy2015SNScrosscorrelations}.
When the two species experience different wall couplings, their spin
dynamics is determined by different diffusion-mode bases. Therefore
mutual spin exchange, which is due to a local coupling (atom-atom
collisions), depends on the mode overlap between these bases.

Here we consider the coupling of alkali-metal spins to noble-gas spins,
such as helium-3, for potential applications in quantum optics \citep{AlkaliNobleEntanglementKatz2020PRL}.
The nuclear spins of noble gases are well protected by the enclosing
complete electronic shells and thus sustain many collisions with other
atoms and with the cell walls. Their lifetime typically reaches minutes
and hours \citep{walker1997SEOPReview,gemmel2010UltraSensitiveMagnetometer,Walker2017He3review}.
In an alkali-metal--noble-gas mixture, the noble gas acts as a buffer
both for itself and for the alkali-metal atoms, so that both species
diffuse, and their collective spin states can be described by our
Bloch-Heisenberg-Langevin model.

As the noble-gas spins do not relax by wall collisions, their lowest-order
diffusion mode $u_{0}^{\mathrm{b}}(\mathbf{r})$ is that associated
with the characteristic (extremely) long life time. Higher-order modes
$u_{n}^{\mathrm{b}}(\mathbf{r})$ decay due to diffusion with typical
rates $\Gamma_{\mathrm{wall}}n^{2}=n^{2}\pi^{2}D/R^{2}$, where $R$
is the length scale of the system. For typical systems, $\Gamma_{\mathrm{wall}}$
is of the order of $(\unit[1\,]{ms})^{-1}-(\unit[1\,]{sec})^{-1}$.
Consequently, to enjoy the long lifetimes of noble-gas spins, one
should employ solely the uniform mode.

The alkali-metal spins couple locally to the noble-gas spins with
a collective rate $J$ via spin-exchange collisions \citep{weakcollisions2019arxiv}.
Unlike the noble-gas spins, the alkali-metal spins are strongly affected
by the cell walls, and consequently their low-order diffusion modes
$u_{m}^{\mathrm{a}}(\mathbf{r})$ are different. This mode mismatch,
between $u_{m}^{\mathrm{a}}(\mathbf{r})$ and $u_{n}^{\mathrm{b}}(\mathbf{r})$,
leads to fractional couplings $c_{mn}J$, where $c_{mn}=\int_{V}d^{3}\mathbf{r}\,u_{m}^{\mathrm{a*}}(\mathbf{r})u_{n}^{\mathrm{b}}(\mathbf{r})$
are the overlap coefficients. In particular, $|c_{m0}|J$ are the
couplings to the uniform (long lived) mode of the noble-gas spins.
Usually, no anti-relaxation coating is used in these experiments,
thus $|c_{m0}|$<1.

Here we demonstrate a calculation for a spherical cell of radius $R$,
for which the radial mode bases $u_{m}^{\mathrm{a}}(\mathbf{r})$
and $u_{n}^{\mathrm{b}}(\mathbf{r})$ and associated decay rates $\Gamma_{\mathrm{a}m}$
and $\Gamma_{\mathrm{b}n}$ are presented in Appendix \ref{sec:solution-of-diffusion-relaxation},
alongside the first $c_{m0}$ values for an uncoated cell (Table \ref{tab:overlap-in-spherical-cell}).
The calculation includes the first $m,n\le70$ modes \footnote{Excess decay and noise due to the modes $m,n\ge70$ are introduced
along the lines of Eq. (S4) in Ref.~\citep{weakcollisions2019arxiv}}. As the initial state, we consider the doubly excited (Fock) state
of the alkali-metal spins $|\psi_{0}\rangle=\frac{1}{\sqrt{2}}(\sum_{m}\alpha_{m}a_{m}^{\dagger})^{2}|0\rangle_{\mathrm{a}}|0\rangle_{\mathrm{b}}=|2\rangle_{\mathrm{a}}|0\rangle_{\mathrm{b}}$,
where $|0\rangle_{\mathrm{a}}|0\rangle_{\mathrm{b}}$ is the vacuum
state with all spins pointing downwards. We take the initial excitation
to be spatially uniform, for which the coefficients $\alpha_{m}=c_{m0}$
satisfy $\sum_{m}\alpha_{m}u_{m}(\mathbf{r})=u_{0}^{\mathrm{b}}(\mathbf{r})=(4\pi R^{3}/3)^{-1/2}$.
We calculate the transfer of this excitation via spin exchange to
the uniform mode $\hat{b}_{0}$ of the noble-gas spins, i.e., to the
state $|0\rangle_{\mathrm{a}}|2\rangle_{\mathrm{b}}=2^{-1/2}(\hat{b}_{0}^{\dagger})^{2}|0\rangle_{\mathrm{a}}|0\rangle_{\mathrm{b}}$.

\begin{figure}
\includegraphics[width=0.95\columnwidth]{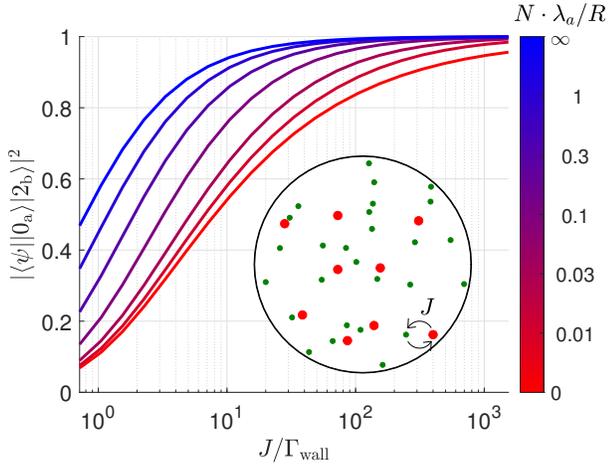}

\caption{Excitation exchange between polarized alkali-metal and noble-gas spins.
Shown is the exchange fidelity of the doubly excited (Fock) states
$|2\rangle_{\mathrm{a}}|0\rangle_{\mathrm{b}}$ and $|0\rangle_{\mathrm{a}}|2\rangle_{\mathrm{b}}$.
We assume a spherical cell containing potassium and helium-3. The
quality $N$ of the wall coating for the alkali metal is varied between
no coating ($N\le1$) and perfect coating ($N\rightarrow\infty$).
The noble-gas spins do not couple to the cell walls. The exchange
fidelity approaches 1 when $J\gg\Gamma_{\mathrm{wall}},\Gamma_{\mathrm{a}}$,
as then the spin exchange is efficient for many diffusion modes; here
$\Gamma_{\mathrm{wall}}$ is the contribution of wall collisions to
the relaxation rate of the alkali-metal spin (i.e., the typical diffusion
rate to the walls), and $\Gamma_{\mathrm{a}}$ is the contribution
of atomic collisions. The calculations are performed for a cell radius
$R=5$ mm and with $1$ atm of helium-3. The diffusion constants are
$D_{\mathrm{a}}=0.35\,\mathrm{cm^{2}/s}$ for the potassium (mean
free path $\lambda_{\mathrm{a}}=50$ nm) so that $\Gamma_{\mathrm{wall}}=\pi^{2}D_{\mathrm{a}}/R^{2}=\unit[1/(70]{\,ms)}$,
and $D_{\mathrm{b}}=0.7\,\mathrm{cm^{2}/s}$ for the helium (mean
free path $\lambda_{\mathrm{b}}=20$ nm). The additional homogeneous
decay of the alkali metal is $\Gamma_{\mathrm{a}}\approx6\,\mathrm{s^{-1}}$
\citep{Happer2010book}. The wall coating plays a significant role,
since for $N\lambda_{\mathrm{a}}/R>1$ (i.e., $N>10^{5}$) the diffusion
modes of the potassium and helium spins match. \label{fig:weak-collisions-exchange-fidelity}}
\end{figure}

Figure \ref{fig:weak-collisions-exchange-fidelity} displays the exchange
fidelity $\mathcal{F}=\max|\langle\psi(t)|\,|0\rangle_{\mathrm{a}}|2\rangle_{\mathrm{b}}|^{2}$
as a function of both spin-exchange rate $J$ and quality of coating
$N$. As $N$ increases, the initial uniform excitation matches better
the lower-order modes of the alkali-metal spins, which couple better
to the uniform modes of the noble-gas spins. Indeed we find that the
exchange fidelity grows with increasing $J$ and $N$.

\section{Discussion\label{sec:Discussion}}

We have presented a fully quantum model, based on a Bloch-Heisenberg-Langevin
formalism, for the effects of diffusion on the collective spin states
in a thermal gas. The model is valid when the atomic mean free path
is much shorter than the apparatus typical dimension. This is often
the case for warm alkali-metal-vapor systems, even when a buffer gas
is not deliberately introduced, as the out-gassing of a spin-preserving
wall coating can lead to mean free paths on the order of millimeters
\citep{sekiguchi2016JapaneseParaffinOutgassing,Hatakeyama2019ParaffinOutgassing2}.

We have mostly focused on highly polarized spin ensembles, typically
used to study nonclassical phenomena that employ the transverse component
of the spin. It is important to note that Eqs.~(\ref{eq:spin-diffusion-equation})
and (\ref{eq:boundary-condition}) hold generally and can be applied
to unpolarized systems as well. For example, the presented analysis
of spin noise spectra holds for unpolarized vapor {[}accounting for
suitable spin statistics in Eq.~(\ref{eq:APN-PSD}) using Eq.~(\ref{eq:Ptilde}){]}
and is thus applicable to nonclassical experiments done in that regime
\citep{Kong2018MitchellAlkaliSEEntanglement}. Our model can also
describe other space-dependent phenomena, such as the dynamics in
the presence of nonuniform driving fields \citep{Xiao2019MultiplexingSqueezedLightDiffusion}.

The presented model agrees with existing mean-field descriptions of
diffusion of atomic spins. It further agrees with models employing
the dissipation-fluctuation theorem to derive the spin noise spectrum
from the decay associated with diffusion. Importantly, it extends
all these models by describing quantum correlations and explicitly
deriving the quantum noise of the Brownian motion. The suggested model
assumes $\lambda\ll R,L$ (Fickian diffusion) and thus does not hold
for the special case of small, low-pressure, coated cells where the
atomic motion is predominantly ballistic ($\lambda\gtrsim R,L$) \citep{Borregaard2016SinglePhotonsOnMotionallyAveragedMicrocellsNcomm,Tang2020DiffusionLowPressure_PRA}.
Nevertheless, it may still provide a qualitative description of the
effect of wall collisions on the uniform spin distribution across
both diffusion regimes \citep{Tang2020DiffusionLowPressure_PRA}.
Our model lays the groundwork for treatments of such systems by considering
non-Markovian motional dynamics.

Our results highlight the multimode nature of the dynamics. As exemplified
for the applications considered in Sec.~\ref{sec:Applications},
one often needs to account for multiple diffusion modes, with the
high-order modes introducing additional quantum noise or reducing
fidelities. As a rule of thumb, if $\varepsilon$ is the allowed infidelity
or excess quantum noise, then one should include the first $\sim\varepsilon^{-1}$
modes in the calculations.

Since thermal motion is inherent to gas-phase systems, our model could
be beneficial to many studies of nonclassical spin gases and particularly
to warm alkali-metal vapors. One such example is a recent demonstration
of transfer of quantum correlations by the diffusion of alkali-metal
atoms between different spatial regions \citep{Xiao2019MultiplexingSqueezedLightDiffusion}.
Other examples involve a single active region, e.g., when spin squeezing
is performed using a small probe beam over a long probing time, with
the goal of coupling efficiently to the uniform diffusion mode in
a coated cell \citep{Polzik2010ReviewRMP,Borregaard2016SinglePhotonsOnMotionallyAveragedMicrocellsNcomm}.
The resulting spatio-temporal dynamics can be described using our
model in order to assess the obtainable degree of squeezing. In particular,
our model predicts that high buffer gas pressure would improve the
lifetime of squeezed states when small probe beams are employed (e.g.,
when using optical cavities or when high probe intensities are needed),
thus encouraging the realization of such experiments.
\begin{acknowledgments}
We thank Eugene Polzik for fruitful discussions and insights. We acknowledge
financial support by a European Research Council starting investigator
grant (Q-PHOTONICS Grant No. 678674), the Israel Science Foundation,
the Pazy Foundation, the Minerva Foundation with funding from the
Federal German Ministry for Education and Research, and the Laboratory
in Memory of Leon and Blacky Broder.
\end{acknowledgments}

\appendix

\section{Diffusion-induced noise\label{sec:diffusion-noise}}

In the main text, we formulate the dynamics of a collective spin operator
as driven from local density fluctuations. For deriving Eq.~(\ref{eq:spin-diffusion-equation}),
we use the Lagrangian version of Eq.$\,$(\ref{eq:Dean-diffusion}),
where the noise is defined for each particle individually 
\begin{equation}
\partial n_{a}/\partial t=D\nabla^{2}n_{a}+\boldsymbol{\nabla}[\boldsymbol{\eta}^{(a)}(t)\sqrt{n_{a}}].\label{eq:Lagrangian-Dean-diffusion}
\end{equation}
Here $\boldsymbol{\eta}^{(a)}(t)$ is a white Gaussian process with
vanishing mean $\langle\boldsymbol{\eta}^{(a)}\rangle_{\mathrm{c}}=\mathbf{0}$
and with correlations $\langle\eta_{i}^{(a)}(t)\eta_{j}^{(a')}(t')\rangle_{\mathrm{c}}=2D\delta_{ij}\delta_{aa'}\delta(t-t')$.
Substituting these into Eq.$\,$(\ref{eq:spin-equation}) provides
the definition for the quantum noise components as 

\begin{equation}
\hat{f}_{\mu}(\mathbf{r},t)=\sum_{a=1}^{N_{\mathrm{a}}}\hat{\mathrm{s}}_{\mu}^{(a)}(t)\boldsymbol{\nabla}[\boldsymbol{\eta}^{(a)}n_{a}(\mathbf{r},t)].\label{eq:Lagnrangian-thermal-motion-noise}
\end{equation}

Following the lines of Ref.~\citep{dean1996langevinDiffusion}, we
consider an alternative, equivalent definition 
\begin{equation}
\hat{f}_{\mu}(\mathbf{r},t)=\boldsymbol{\nabla}[\hat{\mathrm{s}}_{\mu}(\mathbf{r},t)\boldsymbol{\eta}(\mathbf{r},t)\slash\sqrt{n}],\label{eq:Eulerian-thermal-motion-noise}
\end{equation}
as also provided in the main text. According to both definitions,
$\hat{\boldsymbol{f}}$ is a stochastic Gaussian process (linear operations
on a Gaussian process accumulate to a Gaussian process) with a vanishing
mean. Consequently, the equivalence of the two definitions is a result
of the equality of the noise correlations 
\begin{align}
\langle\hat{f}_{\mu}\hat{f}_{\nu}'\rangle_{\mathrm{c}}= & \langle\sum_{i}\nabla_{i}(\sum_{a}\hat{\mathrm{s}}_{\mu}^{(a)}n_{a}\eta_{i}^{(a)})\times\nonumber \\
 & \:\sum_{j}\nabla'_{j}(\sum_{a'}\hat{\mathrm{s}}_{\nu}^{(a')}n_{a'}\eta_{j}^{(a')})\rangle_{\mathrm{c}}\nonumber \\
= & 2D(\boldsymbol{\nabla}\cdot\boldsymbol{\nabla}')(\sum_{a}\hat{\mathrm{s}}_{\mu}^{(a)}\hat{\mathrm{s}}_{\nu}^{(a)}n_{a})\delta(\mathbf{r}-\mathbf{r}')\delta(t-t')\nonumber \\
= & 2D(\boldsymbol{\nabla}\cdot\boldsymbol{\nabla}')\frac{\sum_{aa'}\hat{\mathrm{s}}_{\mu}^{(a)}n_{a}\hat{\mathrm{s}}_{\nu}^{(a')}n_{a'}}{\sum_{a'}n_{a'}}\delta(\mathbf{r}-\mathbf{r}')\delta(t-t')\nonumber \\
= & 2D(\boldsymbol{\nabla}\cdot\boldsymbol{\nabla}')(\hat{\mathrm{s}}_{\mu}\hat{\mathrm{s}}_{\nu}'/\sqrt{nn'})\delta(\mathbf{r}-\mathbf{r}')\delta(t-t')\nonumber \\
= & \langle\sum_{i}\nabla_{i}(\hat{\mathrm{s}}_{\mu}\eta_{i}/\sqrt{n})\times\sum_{j}\nabla_{j}'(\hat{\mathrm{s}}_{\nu}'\eta_{j}'/\sqrt{n'})\rangle_{\mathrm{c}},\label{eq:equivalence-of-eulerian-lagrangian-thermal-motion-quantum-noises}
\end{align}
where we used the identity $n_{a}(\mathbf{r},t)n_{a'}(\mathbf{r},t)=\delta_{aa'}n_{a}(\mathbf{r},t)n_{a'}(\mathbf{r},t)$.
Here and henceforth, we use tags to abbreviate the coordinates $(\mathbf{r}',t')$
for a field, i.e., $F'=F(\mathbf{r}',t')$ and $F=F(\mathbf{r},t)$.

The quantum noise, the commutation relations of which are shown in
Eq.$\,$(\ref{eq:diffusion-noise-commutation-relation}), conserves
the spin commutation relations $[\hat{\mathrm{s}}_{\mu}(\mathbf{r},t),\hat{\mathrm{s}}_{\nu}(\mathbf{r}',t)]=i\epsilon_{\xi\mu\nu}\hat{\mathrm{s}}_{\xi}\delta(\mathbf{r}-\mathbf{r}')$.
This can be seen from

\begin{align}
\langle[\hat{\mathrm{s}}_{\mu}(\mathbf{r},t+dt),\hat{\mathrm{s}}_{\nu}(\mathbf{r}',t+dt)]-[\hat{\mathrm{s}}_{\mu}(\mathbf{r},t),\hat{\mathrm{s}}_{\nu}(\mathbf{r}',t)]\rangle_{\mathrm{c}}\\
=i\epsilon_{\xi\mu\nu}\delta(\mathbf{r}-\mathbf{r}')\langle\hat{\mathrm{s}}_{\xi}(\mathbf{r},t+dt)-\hat{\mathrm{s}}_{\xi}(\mathbf{r},t)\rangle_{\mathrm{c}}\nonumber 
\end{align}
and then
\begin{align}
i\epsilon_{\xi\mu\nu}D(\boldsymbol{\nabla}+\boldsymbol{\nabla}')^{2}[\hat{\mathrm{s}}_{\xi}\delta(\mathbf{r}-\mathbf{r}')]dt\label{eq:conservation-of-spin-commutation-relations}\\
=i\epsilon_{\xi\mu\nu}D(\nabla^{2}\hat{\mathrm{s}}_{\xi})\delta(\mathbf{r}-\mathbf{r}')dt,\nonumber 
\end{align}
where the last equality stems from $(\boldsymbol{\nabla}+\boldsymbol{\nabla}')\delta(\mathbf{r}-\mathbf{r}')=0$.

In Sec.~\ref{sec:Polarized-ensemb}, we focus on highly polarized
ensembles, where the dynamics is described by the bosonic annihilation
operator $\hat{a}$, under the Holstein-Primakoff approximation. Under
these conditions, the thermal noise operating on the bosonic excitations
becomes $\hat{f}=\boldsymbol{\nabla}(\hat{a}\boldsymbol{\eta}/\sqrt{n})$.
In addition, the same conditions ensure that $\hat{a}^{\dagger}\hat{a}=0$,
$\hat{a}(\mathbf{r},t)\hat{a}^{\dagger}(\mathbf{r}',t)=\delta(\mathbf{r}-\mathbf{r}')$,
and $\hat{a}(\mathbf{r},t)\hat{a}^{\dagger}(\mathbf{r}',t)\delta(\mathbf{r}-\mathbf{r}')=n\delta(\mathbf{r}-\mathbf{r}')$,
thus providing 
\begin{align}
\langle\hat{f}\hat{f}'^{\dagger}\rangle_{\mathrm{c}} & =\langle\boldsymbol{\nabla}\hat{a}\boldsymbol{\eta}/\sqrt{n}\,\boldsymbol{\nabla}'(\hat{a}^{\dagger})'\boldsymbol{\eta}'/\sqrt{n'}\rangle_{\mathrm{c}}\nonumber \\
 & =-2D\nabla^{2}\delta(\mathbf{r}-\mathbf{r}')\delta(t-t'),
\end{align}
and $\langle\hat{f}^{\dagger}\hat{f}'\rangle_{\mathrm{c}}=0$. Therefore,
the noise becomes a vacuum noise and conserves the commutation relations
of the bosonic operators. We denote the correlations of the diffusion
noise in the bulk as $C(\mathbf{r},\mathbf{r}')=-2D\nabla^{2}\delta(\mathbf{r}-\mathbf{r}')$.

\section{Model for wall coupling\label{sec:wall-coupling-toy}}

We adopt a simplified model for describing the scattering of atoms
off the cell walls. The model assumes that the wall coupling is stochastic
and Markovian, thus resulting in an exponential decay of the scattered
spin, and that the noise due to diffusion in the bulk vanishes within
a thin boundary layer at the wall. This leads to the scattering described
by Eq.$\,$(\ref{eq:wall-scattering}). The accompanying noise processes
for atoms $a$ and $a'$ satisfy the relations 
\begin{equation}
[\hat{w}_{a}^{\mu}(t),\hat{w}_{a'}^{\nu}(t')]=i\epsilon_{\mu\nu\xi}e^{-1/N}(1-e^{-1/N})\hat{\mathrm{s}}_{\xi}^{(a)}\delta_{aa'}\frac{\varpi\delta(t-t')}{\bar{v}},\label{eq:lagrangian-wall-scattering-fluctuation}
\end{equation}
where $\mu,\nu=x,y,z$. Here $\varpi=(e^{1/N}-1)^{-1}\lambda/3$ is
the effective correlation distance of the wall-scattering noise, defined
such that the commutation relations of the spin operators are conserved
for all diffusion modes, i.e., for the entire cell (bulk and boundary).
It changes monotonically from $\varpi=e^{-1/N}\lambda/3$ for spin-destructing
walls ($N\ll1)$ to $\varpi=N\lambda/3$ for spin-preserving walls
($N\gg1$).

The continuous operator $\hat{\boldsymbol{w}}(\mathbf{r},t)$ used
in the main text to describe the noise due to interactions with the
cell walls is defined as $\hat{\boldsymbol{w}}=\sum_{a}\hat{\boldsymbol{w}}_{a}n_{a}$.
It is the analog of $\hat{\boldsymbol{w}}_{a}(t)$, like $\hat{\mathbf{s}}$
is to $\hat{\mathbf{s}}_{a}$. It vanishes for positions $\mathbf{r}$
the distance of which from the boundary is larger than $\varpi$,
and its commutation relations are

\begin{align}
\langle[\hat{w}^{\mu},\hat{w}'^{\nu}]\rangle_{\mathrm{c}}= & i\epsilon_{\mu\nu\xi}e^{-1/N}(1-e^{-1/N})\varpi/\bar{v}\,\times\nonumber \\
 & \hat{\mathrm{s}}_{\xi}(\mathbf{r},t)\delta(\mathbf{r}-\mathbf{r}')\delta(t-t').\label{eq:eulerian-wall-noise-commutation}
\end{align}
The last expression is defined only for coordinates $\mathbf{r}$
and $\mathbf{r}'$ on the cell boundary and vanishes elsewhere. As
an example, for a rectangular cell with a wall at $x=L/2$, we shall
define coordinates on the boundary $\mathbf{r}_{\perp}=y\hat{\mathbf{y}}+z\hat{\mathbf{z}}$
and substitute $\delta(\mathbf{r}-\mathbf{r}')=\frac{1}{\varpi}\delta(y-y')\delta(z-z')$
at $x=x'=L/2$. For a spherical cell with a wall at $|\mathbf{r}|=R$,
we use $\delta(\mathbf{r}-\mathbf{r}')=\frac{1}{\varpi}\frac{\delta(\Omega-\Omega')}{R^{2}}$,
where $\Omega$ is the angular position of coordinate $\mathbf{r}$.

Using $\hat{\boldsymbol{w}}(\mathbf{r},t)$, the scattering matrix
for the spin-density operator becomes $\mathcal{S}\hat{\mathbf{s}}=e^{-1/N}\cdot\hat{\mathbf{s}}+\hat{\boldsymbol{w}}$.
We write Eq.~(\ref{eq:boundary-condition}) for the spin density
operator using the noise field $\hat{\boldsymbol{w}}$. In addition,
$\hat{\boldsymbol{w}}$ is defined only on the boundary, such that
$\left.(\hat{\mathbf{n}}\cdot\boldsymbol{\nabla})\hat{\boldsymbol{w}}\right|_{\text{boundary}}\propto\delta^{\prime}(0)$
and therefore vanishes.

Finally, under the Holstein-Primakoff approximation, we use Eq.$\,$(\ref{eq:eulerian-wall-noise-commutation})
to find the noise operating on $\hat{a}$ due to wall scattering.
The operator $\hat{w}=\hat{w}_{-}/\sqrt{2|\mathrm{s}_{z}|}$ becomes
a vacuum noise, satisfying $\langle\hat{w}^{\dagger}\hat{w}'\rangle_{\mathrm{c}}=0$,
and for $\mathbf{r}$ and $\mathbf{r}'$ on the cell boundary, 
\begin{equation}
\langle\hat{w}\hat{w}'^{\dagger}\rangle_{\mathrm{c}}=2e^{-1/N}(1-e^{-1/N})\varpi/\bar{v}\delta(\mathbf{r}-\mathbf{r}')\delta(t-t').\label{eq:wall-vacuum-noise-commutation}
\end{equation}

Considering a general spin distribution, the noise due to the walls
exists only in a volume of order $\varpi S$, where $S$ is the cell
surface area, while the noise due to diffusion in the bulk exists
in the entire volume $V$. The ratio of the two scales as $\langle\hat{w}\hat{w}'^{\dagger})\rangle_{\mathrm{c}}/\langle\hat{f}\hat{f}'^{\dagger})\rangle_{\mathrm{c}}\propto\varpi S/V\propto\lambda/R$,
where $R$ is the typical dimension of the cell. Consequently, in
our considered diffusive regime $\lambda\ll R$, the diffusion noise
dominates over that of the wall scattering for nonuniform spin distributions.

\section{Solving the diffusion-relaxation Bloch-Heisenberg-Langevin equations\label{sec:solution-of-diffusion-relaxation}}

The diffusion-relaxation equation in the Bloch-Heisenberg-Langevin
formalism, in the limit of a highly polarized spin gas, is presented
in Sec.~\ref{sec:Polarized-ensemb}. Here we first solve Eqs.~(\ref{eq:HP heisenberg langevin})
and (\ref{eq:HP boundary condition}) for a simplified 1D case by
following the method described in the main text. We provide explicit
expressions for the mode-specific noise sources due to motion in the
bulk and at the boundary. Finally, we provide tabulated solutions
for the three-dimensional cases of rectangular, cylindrical, and spherical
cells.

Consider a 1D cell with a single spatial coordinate $-L/2\le x\le L/2$.
The functions $u_{k}(x)$ that solve the Helmholtz equation $\partial^{2}u_{k}/\partial x^{2}+k^{2}u_{k}=0$
are the relaxation-diffusion modes, where the decay rates $\Gamma$
introduced in the main text are $\Gamma=Dk^{2}$. These solutions
are $u_{k}^{+}=A_{k}^{+}\cos(k^{+}x)$ and $u_{k}^{-}=A_{k}^{-}\sin(k^{-}x)$,
composing symmetric and anti-symmetric modes. The annihilation operator
decomposes into a superposition $\hat{a}(x,t)=\sum_{k,\pm}\hat{a}_{k}^{\pm}(t)u_{k}^{\pm}(x)$.
To further simplify the example, we take only symmetric spin distribution
and symmetric noise into consideration, i.e., we keep only the modes
$u_{k}^{+}$ and omit the ``$+$'' superscript. Note that a physical
noise is random and generally has no defined symmetry, but it can
be decomposed into components with well-defined symmetry.

The bulk diffusion equation becomes $\partial\hat{a}_{k}/\partial t=i[\mathcal{H},\hat{a}_{k}]-Dk^{2}\hat{a}_{k}+\int_{-L/2}^{L/2}\hat{f}u_{k}dx$.
We break the boundary equation into a homogeneous part, where the
noise is omitted, and an inhomogeneous part, which includes the noise.
The former can be decomposed into the different modes and is simplified
to the algebraic equation $\cot(kL/2)=2\frac{1+e^{-1/N}}{1-e^{-1/N}}\lambda k$
\footnote{Here in one dimension, we use the relation $D=\lambda\bar{v}$ between
the diffusion coefficient and the kinetic parameters. In three dimensions,
this becomes $D=\lambda\bar{v}/3$, such that $\cot\left(kL/2\right)=\frac{2}{3}\frac{1+e^{-1/N}}{1-e^{-1/N}}\lambda k$.
$\varpi$ has similar dimensionality dependence. }.

\begin{figure}
\includegraphics[width=0.9\columnwidth]{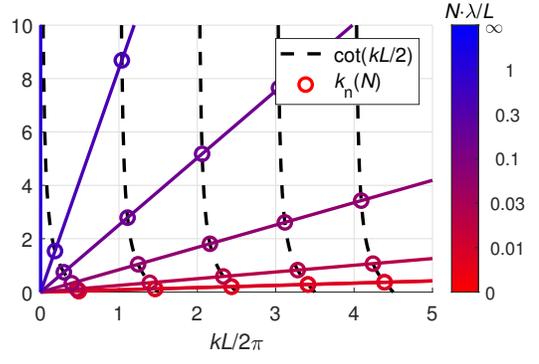}

\caption{Graphical solutions for the Robin boundary condition. Here we solve
the 1D equations for a system of length $L=\unit[1]{cm}$ and mean
free path of $\lambda=\unit[0.5]{\mu m}$ (characteristic of $100$
Torr of buffer gas) and for different values of $N$.\label{fig:graphical-solution-for-Robin-condition}}
\end{figure}

For general values of $N$, this is a Robin boundary condition, which
can be solved numerically or graphically as presented in Fig.~\ref{fig:graphical-solution-for-Robin-condition}.
The discrete solutions $k_{n}$ define a complete and orthonormal
set of discrete modes $u_{n}=A_{n}\cos(k_{n}x)$, spanning all symmetric
spin distributions in the 1D cell, and $\int_{-L/2}^{L/2}u_{m}^{\ast}u_{n}dx=\delta_{nm}$.
These provide the discrete decay rates $\Gamma_{n}$.

For example, in the Dirichlet case of destructive walls ($N\lambda/L\ll1$),
$k_{n}=(2n+1)\pi/L$. The annihilation operators of the various modes
are $\hat{a}_{n}(t)=\int_{-L/2}^{L/2}u_{n}^{\ast}(x)\hat{a}(x,t)dx$,
and the noise operators are $\hat{f}_{n}(t)=\int_{-L/2}^{L/2}u_{n}^{\ast}(x)\hat{f}(x,t)dx$
and $\hat{w}(t)=[\hat{w}(L/2,t)+\hat{w}(-L/2,t)]/2$.

The treatment of $\hat{f}_{n}$ as a bulk source term operating on
independent modes is a common technique \citep{SteckNotes}. It differs,
however, from the treatment of the noise at the boundaries. We deal
with this term by defining auxiliary fields 
\begin{equation}
\hat{a}(x,t)=\hat{p}(x,t)+\sum_{n}\hat{h}_{n}(t)u_{n}(x),
\end{equation}
as we desire to use $\hat{p}(x,t)$ to imbue the wall noise as a source
acting on the modes $\hat{a}_{n}$, while $\hat{h}_{n}$ solves the
homogeneous equations in the absence of wall-induced fluctuations.
Therefore $\hat{p}(x,t)$ is defined such that $\nabla^{2}\hat{p}(x,t)=0$.
\begin{center}
\begin{table*}
\begin{centering}
\begin{tabular}{|>{\centering}p{0.12\paperwidth}||>{\centering}p{0.23\paperwidth}|>{\centering}p{0.2\paperwidth}|>{\centering}p{0.24\paperwidth}|}
\hline 
\centering{}cell shape & \centering{}rectangular & \centering{}cylindrical & \centering{}spherical\tabularnewline
\hline 
\hline 
\centering{}symmetry & $\begin{array}{c}
\text{symmetric: }(+)\\
\text{anti-symmetric: }(-)
\end{array}$ & \centering{}angular: $n$ & \centering{}spherical: $\ell,p$\tabularnewline
\hline 
\centering{}coordinate range & \centering{}$-L/2\leq x\leq L/2$ & $\begin{array}{c}
0\leq\rho\leq R\\
0\leq\varphi\leq2\pi
\end{array}$ & $\begin{array}{c}
0\leq r\leq R\\
0\leq\theta\leq\pi\\
0\leq\varphi\leq2\pi
\end{array}$\tabularnewline
\hline 
\centering{}boundary equation & $\begin{array}{c}
\cot(k_{n}^{+}L/2)=\frac{2}{3}\frac{1+e^{-1/N}}{1-e^{-1/N}}\lambda k_{n}^{+}\\
-\tan(k_{n}^{-}L/2)=\frac{2}{3}\frac{1+e^{-1/N}}{1-e^{-1/N}}\lambda k_{n}^{-}
\end{array}$ & \centering{}$-\frac{J_{n}(k_{\nu n}R)}{J_{n}^{\prime}(k_{\nu n}R)}=\frac{2}{3}\frac{1+e^{-1/N}}{1-e^{-1/N}}\lambda k_{\nu n}$ & \centering{}$-\frac{j_{\ell}(k_{n\ell}R)}{j_{\ell}^{\prime}(k_{n\ell}R)}=\frac{2}{3}\frac{1+e^{-1/N}}{1-e^{-1/N}}\lambda k_{n\ell}$\tabularnewline
\hline 
\centering{}$u_{n}(\mathbf{r})$ & $\begin{array}{c}
u_{n}^{+}(x)=A_{n}^{+}\cos(k_{n}^{+}x)\\
u_{n}^{-}(x)=A_{n}^{-}\sin(k_{n}^{-}x)
\end{array}$ & \centering{}$u_{\nu n}(\rho,\varphi)=A_{\nu n}J_{n}(k_{\nu n}\rho)e^{in\varphi}$ & \centering{}$u_{n\ell p}(r,\theta,\varphi)=A_{n\ell p}j_{\ell}(k_{n\ell}r)Y_{\ell p}(\theta,\varphi)$\tabularnewline
\hline 
\end{tabular}
\par\end{centering}
\caption{Solutions of the diffusion-relaxation modes for rectangular, cylindrical,
and spherical cells. $J_{n}(x)$ is the $n$th Bessel function of
the first kind, $j_{\ell}(x)$ is the $\ell$th spherical Bessel function
of the first kind, and $Y_{\ell p}(\theta,\varphi)$ are the spherical
harmonics. The decay rates satisfy $\Gamma_{n}=Dk_{n}^{2}$.\label{tab:Diffusion-mode-solutions}}
\end{table*}
\par\end{center}

In our 1D symmetric case, $\hat{p}(x,t)=\hat{p}(t)$ is uniform. Writing
the full boundary equation for $\hat{a}$ provides $\hat{p}(t)=\hat{w}(t)/(1-e^{-1/N})$.
We decompose $\hat{p}(t)$ into the modes to obtain $\hat{p}_{n}(t)=\int_{-L/2}^{L/2}\hat{p}(t)u_{n}(x)dx=2A_{n}\sin(k_{n}L/2)\hat{p}(t)/k_{n}$.
Substituting this in Eq.$\,$(\ref{eq:HP heisenberg langevin}) provides
the equation for the homogeneous mode operators $\hat{h}_{n}$.

In the case of a magnetic Zeeman Hamiltonian $\mathcal{H}=i\omega_{0}\hat{\mathrm{S}}_{z}$,
we find 
\begin{equation}
\partial\hat{h}_{n}/\partial t=-i\omega_{0}\hat{h}_{n}-\Gamma_{n}\hat{h}_{n}+\hat{f}_{n}-i\omega_{0}\hat{p}_{n}-\partial\hat{p}_{n}/\partial t,\label{eq:boundaryless-diffusion-eqn-eqn}
\end{equation}
the solutions of which are 
\begin{equation}
\begin{array}{cc}
\hat{h}_{n}= & e^{-(i\omega_{0}+\Gamma_{n})t}\hat{h}_{n}\left(0\right)+\qquad\qquad\qquad\qquad\qquad\qquad\qquad\\
 & \int_{0}^{t}e^{-(i\omega_{0}+\Gamma_{n})(t-\tau)}(\hat{f}_{n}(\tau)-(i\omega_{0}+\tfrac{\partial}{\partial\tau})\hat{p}_{n}(\tau))d\tau.
\end{array}\label{eq:boundaryless-diffusion-eqn-solution}
\end{equation}
Substituting into $\hat{a}_{n}(t)=\hat{p}_{n}(t)+\hat{h}_{n}(t)$
and differentiating with respect to $t$ provides the evolution of
the annihilation operators of the spin modes 
\begin{equation}
\partial\hat{a}_{n}/\partial t=-(i\omega_{0}+\Gamma_{n})\hat{a}_{n}+\hat{f}_{n}+\hat{f}_{n}^{\mathrm{w}},\label{eq:mode-operator-diffusion-eqn}
\end{equation}
where 
\begin{equation}
\hat{f}_{n}^{\mathrm{w}}=\Gamma_{n}\int_{-L/2}^{L/2}u_{n}^{\ast}(x)\hat{p}(x,t)dx=\frac{2A_{n}\Gamma_{n}\sin(k_{n}L/2)}{(1-e^{-1/N})k_{n}}\hat{w}
\end{equation}
is the quantum noise due to wall collisions. Finally, we can combine
the two noise terms and obtain the total, mode-specific, noise operator
\begin{equation}
\hat{\mathcal{W}}_{n}=\int_{0}^{t}e^{-[i\omega_{0}+D(k_{n}^{\pm})^{2}](t-\tau)}[\hat{f}_{n}(\tau)+\hat{f}_{n}^{\mathrm{w}}(\tau)]d\tau,\label{eq:mode-noise-process}
\end{equation}
appearing in Eq.~(\ref{eq:eigenmode-time-evolution}).

Under the influence of the noise sources $\hat{\mathcal{W}}_{n}$
and the dissipation $\Gamma_{n}$, the spin operators of the diffusion
modes obey the fluctuation-dissipation theorem, and their commutation
relations are conserved, resulting from $\langle(\hat{f}_{n'}+\hat{f}_{n'}^{\mathrm{w}})(\hat{f}_{n}^{\dagger}+\hat{f}_{n}^{\mathrm{w}\dagger})\rangle_{\mathrm{c}}=2\Gamma_{n}\delta_{n'n}\delta(t-t')$
and $\langle(\hat{f}_{n'}^{\dagger}+\hat{f}_{n'}^{\mathrm{w}\dagger})(\hat{f}_{n}+\hat{f}_{n}^{\mathrm{w}})\rangle_{\mathrm{c}}=0$.
Note that the conservation of local commutation relations is already
presented in Appendices \ref{sec:diffusion-noise} and \ref{sec:wall-coupling-toy}
(where $\hat{f}$ applies for the bulk and $\hat{w}$ for the boundary)
without the mode decomposition. Notably, however, it also holds for
the nonlocal (diffusion) modes.

For completeness, we provide in Table \ref{tab:Diffusion-mode-solutions}
the diffusion-relaxation modes for rectangular, cylindrical, and spherical
cells. Various applications, such as those involving collisional (local)
coupling between two spin ensembles, also require the overlap coefficients
$c_{mn}=\int_{V}d^{3}\mathbf{r}A_{m}^{\ast}(\mathbf{r})B_{n}(\mathbf{r})$
between diffusion modes $A_{m}(\mathbf{r})$ and $B_{n}(\mathbf{r})$.
These are presented in Table \ref{tab:overlap-in-spherical-cell}
for spherically-symmetric modes, where $A_{m}(\mathbf{r})$ are modes
for highly destructive walls ($N\lesssim1$), and $B_{n}(\mathbf{r})$
are for inert walls ($N\gg L/\lambda$). These conditions are typical
for a mixture of alkali-metal vapor and noble gas, as discussed in
section$\,$\ref{sec:Applications}.
\begin{center}
\begin{table}
\begin{tabular}{|c|c|c|c|c|c|}
\hline 
$c_{mn}$ & $n=0$ & $n=1$ & $n=2$ & $n=3$ & $n=4$\tabularnewline
\hline 
$m=0$ & 0.780 & 0.609 & -0.126 & 0.058 & -0.033\tabularnewline
\hline 
$m=1$ & -0.390 & 0.652 & 0.622 & -0.158 & 0.079\tabularnewline
\hline 
$m=2$ & 0.260 & -0.274 & 0.647 & 0.627 & -0.173\tabularnewline
\hline 
$m=3$ & -0.195 & 0.182 & -0.256 & 0.644 & 0.629\tabularnewline
\hline 
$m=4$ & 0.156 & -0.139 & 0.1680 & -0.246 & 0.643\tabularnewline
\hline 
\end{tabular}

\caption{Overlap coefficients, $c_{mn}=\int_{V}d^{3}\mathbf{r}A_{m}^{\ast}(\mathbf{r})B_{n}(\mathbf{r})$,
of the first five spherically symmetric modes, i.e., $\ell=p=0$.
We take $A_{m}(\mathbf{r})$ to be the diffusion modes of a spherical
cell with radius $R=1$ and destructive walls, and $B_{n}(\mathbf{r})$
to be the modes in the same cell but with spin-conserving walls.\label{tab:overlap-in-spherical-cell}}
\end{table}
\par\end{center}

\section{Faraday rotation measurement setup\label{sec:Faraday-rotation-measurement}}

In Sec.~\ref{sec:Applications}, we consider two experimental setups
where the transverse component of a polarized spin ensemble is measured
by means of the Faraday rotation. This scheme is common in alkali-metal
spin measurements \citep{Braginsky1996QNDreview,Julsgaard2001PolzikEntanglement,Kong2018MitchellAlkaliSEEntanglement,AppeltHapper1998SEOPtheoryPRA}.
As illustrated in Fig.~\ref{fig:APN-PSD}a, we consider a cylindrical
cell with radius $R$ and length $L$, with the cylinder axis along
$\hat{\mathbf{x}}$. The spins are polarized along $\hat{\mathbf{z}}$,
parallel to an external applied magnetic field $\mathbf{B}=B\hat{\mathbf{z}}$.
We use $\rho$ and $\varphi$ as the cylindrical coordinates, and
$x$ as the axial coordinate.

A linearly polarized probe beam travels along $\hat{\mathbf{x}}$
with a Gaussian intensity profile $I_{\mathrm{G}}(\mathbf{r})=I_{0}\exp(-2\rho^{2}/w_{0}^{2})$,
where $w_{0}$ is the beam waist radius. We assume a negligible beam
divergence within the cell and require the normalization $\int_{V}I_{\mathrm{G}}^{2}(\mathbf{r})d^{3}\mathbf{r}=1$,
so that $(I_{0})^{-2}=\pi Lw_{0}^{2}(1-e^{-4R^{2}/w_{0}^{2}})/4$.
The probe frequency is detuned from the atomic transition, such that
the probe is not depleted and does not induce additional spin decay.

The linear polarization of the probe rotates due to the Faraday effect,
with the rotation angle proportional to the spin projection along
the beam propagation direction. Therefore, measurement of the rotation
angle provides a measurement of $\hat{\mathrm{s}}_{x}$ weighted by
its overlap with the beam profile. Precisely, the operator $\hat{x}_{\mathrm{G}}(t)=\int_{V}d^{3}\mathbf{r}I_{\mathrm{G}}(\mathbf{r})\hat{x}(\mathbf{r},t)$,
where $\hat{x}(\mathbf{r},t)=[\hat{a}(\mathbf{r},t)+\hat{a}^{\dagger}(\mathbf{r},t)]/2$,
is measured in this scheme \citep{Polzik2010ReviewRMP}.

We identify the atomic diffusion modes in the cylindrical cell as
$u_{n}(\mathbf{r})$. Note that in Table~\ref{tab:Diffusion-mode-solutions}
the modes require several labels, which we replace here with a single
label $n$ for brevity. We decompose the spin operator and the probe
intensity profile using the modes $\hat{x}(\mathbf{r},t)=\sum_{n}\hat{x}_{n}(t)u_{n}(\mathbf{r})$
and $I_{\mathrm{G}}(\mathbf{r})=\sum_{n}I_{n}^{(\mathrm{G})}u_{n}(\mathbf{r})$,
where $\hat{x}_{n}(t)=(\hat{a}_{n}+\hat{a}_{n}^{\dagger})/2$ and
$I_{n}^{(\mathrm{G})}=\int_{V}d^{3}\mathbf{r}I_{\mathrm{G}}(\mathbf{r})u_{n}^{\ast}(\mathbf{r})$.
Using these, we express the measured spin operator as $\hat{x}_{\mathrm{G}}(t)=\sum_{n}I_{n}^{(\mathrm{G})}\hat{x}_{n}(t)$.

We calculate the spin noise spectrum from its formal definition 
\begin{equation}
S_{xx}(f)=\lim_{T\rightarrow\infty}\frac{1}{T}\int_{0}^{T}\int_{0}^{T}\hat{x}_{\mathrm{G}}(\tau)\hat{x}_{\mathrm{G}}(\tau')e^{2\pi if(\tau-\tau')}d\tau d\tau'
\end{equation}
utilizing the temporal evolution of the modes as given by Eq.$\,$(\ref{eq:eigenmode-time-evolution}),
and including the noise properties $\hat{\mathcal{W}}_{n'}^{\dagger}(t)\hat{\mathcal{W}}_{n}(t)=0$
and $\hat{\mathcal{W}}_{n'}(t)\hat{\mathcal{W}}_{n}^{\dagger}(t)=(1-e^{-2\Gamma_{n}t})\delta_{n'n}$
derived from Appendix$\,$\ref{sec:solution-of-diffusion-relaxation}.
The spin noise spectral density appearing in Eq.$\,$(\ref{eq:APN-PSD})
holds for both polarized and unpolarized ensembles, with 

\begin{equation}
\tilde{P}=\begin{cases}
1 & \mathrm{polarized}\\
2(S+1)/3 & \mathrm{unpolarized}
\end{cases}\label{eq:Ptilde}
\end{equation}
where $S$ is the single-particle spin magnitude.

For the considered geometry, the standard quantum limit is $\langle\hat{\mathrm{S}}_{x}^{2}\rangle\ge N_{\text{beam}}/4=\frac{nV_{\text{beam}}}{4}\frac{[1-\exp(-2R^{2}/w_{0}^{2})]^{2}}{1-\exp(-4R^{2}/w_{0}^{2})}$,
where $N_{\text{beam}}=n[\int_{V}I_{\mathrm{G}}(\mathbf{r})d^{3}r]^{2}\slash\int_{V}I_{\mathrm{G}}^{2}(\mathbf{r})d^{3}r$
is the number of atoms in the beam, and $V_{\text{beam}}=\pi Lw_{0}^{2}$
is the beam volume \citep{Shah2010highBWmagnetometryAndAnalysis}.

\bibliographystyle{\string"C:/my programs/MiKTeX 2.9/bibtex/bst/revtex/apsrev4-2\string"}
\bibliography{diffusion_bib}

\end{document}